# Challenges of Next-Generation Wireless Sensor Networks and its impact on Society

Neelam Srivastava

**Abstract**—Wireless Sensor Networks (WSNs) have gained worldwide attention in recent years, particularly with the proliferation in Micro-Electro-Mechanical Systems (MEMS) technology which has facilitated the development of smart sensors. The paper discusses about classification of WSN and challenges of the Next Generation WSN. One of the major challenges of Next Generation WSN is reduction of power consumption. The two approaches are discussed: Ultra-Low-Power Networks and Energy Harvesting. The paper also discusses about some major applications as designing low cost secured Intelligent Buildings, In-Home Health care and Agriculture.

**Index Terms**—Wireless Sensors Networks, Energy Harvesting, Intelligent Building, In-Home Health care, Agriculture.

——————————— ◆ ———————————

## 1 INTRODUCTION

The first wireless trend known as Wireless voice networks emerged in the 1980s. By 1999, wireless data networks had begun. Today, we are entering the third wireless revolution. It is also known as the *Internet of Things*, the third wave is utilizing wireless sense and control technology to bridge the gap between the physical world of humans and the virtual world of electronics. The dream is to automatically monitor and respond to forest fires, avalanches, hurricanes, faults in country wide utility equipment, traffic, hospitals and much more over wide areas and with billions of sensors. It will become possible due to the development of Wireless Sensor Networks (WSN) otherwise known as Ubiquitous Sensor Networks (USN).

Wireless sensor networks (WSNs) have gained worldwide attention in recent years, particularly with the proliferation in Micro-Electro-Mechanical Systems (MEMS) technology which has facilitated the development of smart sensors. These sensors are small, with limited processing and computing resources, and they are inexpensive compared to traditional sensors. These sensor nodes can sense, measure, and gather information from the environment and, based on some local decision process, they can transmit the sensed data to the user. Smart sensor nodes are low power devices equipped with one or more sensors, a processor, memory, a power supply, a radio, and an actuator [1].A variety of mechanical, thermal, biological, chemical, optical, and magnetic sensors may be attached to the sensor node to measure properties of the environment. Since the sensor nodes have limited memory and are typically deployed in difficult-to-access locations, a radio is implemented for wireless communication to transfer the data to a base station (e.g., a laptop, a personal handheld device, or an access point to a fixed infrastructure)[2]. Battery is the main power source in a sensor node. Secondary power supply that harvests power from the environment such as solar panels may be added to the node depending on the appropriateness of the environment where the sensor will be deployed. Depending on the application and the type of sensors used, actuators may be incorporated in the sensors.

A WSN typically has little or no infrastructure. It consists of a number of sensor nodes (few tens to thousands) working together to monitor a region to obtain data about the environment [3]. There are two types of WSNs: Structured and Unstructured. An unstructured WSN is one that contains a dense collection of sensor nodes. Sensor nodes may be deployed in an ad hoc manner into the field. Once deployed, the network is left unattended to perform monitoring and reporting functions. In an unstructured WSN, network maintenance such as managing connectivity and detecting failures is difficult since there are so many nodes. In a structured WSN, all or some of the sensor nodes are deployed in a pre-planned manner The advantage of a structured network is that fewer nodes can be deployed with lower network maintenance and management cost. Fewer nodes can be deployed now since nodes are placed at specific locations to provide coverage while ad hoc deployment can have uncovered regions.

## 2 TYPES OF SENSOR NETWORKS

Current WSNs are deployed on land, underground, and underwater. Depending on the environment, a sensor network faces different challenges and constraints[4].
There are five types of WSNs:

————————————————
- *Author is with Department of Electronics Engineering, Institute of Engineering & Technology. U. P. Technical University, Lucknow, Uttar Pradesh India.*

- Terrestrial WSN,
- Underground WSN,
- Underwater WSN,
- Multi-media WSN, and
- Mobile WSN

Terrestrial WSNs typically consist of hundreds to thousands of inexpensive wireless sensor nodes deployed in a given area, either in an ad hoc or in a pre-planned manner. In ad hoc deployment, sensor nodes can be dropped from a plane and randomly placed into the target area. In pre-planned deployment, there is grid placement, optimal placement, 2-d and 3-d placement and models. Here energy can be conserved with multi-hop optimal routing, short transmission range, in-network data aggregation, eliminating data redundancy, minimizing delays, and using low duty-cycle operations[5].

Underground WSNs and consist of a number of sensor nodes buried underground or in a cave or mine used to monitor underground conditions. **Additional sink nodes are located above ground to relay information from the sensor nodes to the base station**. The underground environment makes wireless communication a challenge due to signal losses and high levels of attenuation. Unlike terrestrial WSNs, the deployment of an underground WSN requires careful planning and energy and cost considerations. Like terrestrial WSN, underground sensor nodes are equipped with a limited battery power and once deployed into the ground, it is difficult to recharge or replace a sensor node's battery. As before, a key objective is to conserve energy in order to increase the lifetime of network which can be achieved by implementing efficient communication protocol.

Underwater WSNs consist of a number of sensor nodes and vehicles deployed underwater. As opposite to terrestrial WSNs, underwater sensor nodes are more expensive and fewer sensor nodes are deployed. Autonomous underwater vehicles are used for exploration or gathering data from sensor nodes.[5] Compared to a dense deployment of sensor nodes in a terrestrial WSN, a sparse deployment of sensor nodes is placed underwater. Typical underwater wireless communications are established through transmission of **acoustic waves**. A challenge in underwater ac communication is the **limited bandwidth**, **long propagation delay**, and **signal fading** issue. Another challenge is sensor node failure due to environmental conditions. Underwater sensor nodes must be able to self-configure and adapt to harsh ocean environment. The issue of energy conservation for underwater WSNs involves developing efficient underwater communication and networking techniques.

Multi-Media WSN have been proposed to enable monitoring and tracking of events in the form of multi-media such as video, audio, and imaging. Multi-media WSNs consist of a number of low cost sensor nodes equipped with cameras and microphones. These sensor nodes interconnect with each other over a wireless connection for data retrieval, process, correlation, and compression. Multi-media sensor nodes are deployed in a pre-planned manner into the environment to guarantee coverage. Challenges in multi-media WSN include **high bandwidth demand**, **high energy consumption**, **quality of service** (QoS) provisioning, **data processing** and **compressing** techniques, and cross-layer design. Multi-media content such as a video stream requires high bandwidth in order for the content to be delivered. As a result, high data rate leads to high energy consumption. Transmission techniques that support high bandwidth and low energy consumption have to be developed. QoS provisioning is a challenging task in a multi-media WSN due to the variable delay and variable channel capacity.

Mobile WSNs consist of a collection of sensor nodes that can move on their own and interact with the physical environment. Mobile nodes have the ability sense, compute, and communicate like static nodes. A key difference is mobile nodes have the ability to reposition and organize itself in the network. A mobile WSN can start off with some initial deployment and nodes can then spread out to gather information. Information gathered by a mobile node can be communicated to another mobile node when they are within range of each other. Another key difference is data distribution. In a static WSN, data can be distributed using fixed routing or flooding while dynamic routing is used in a mobile WSN. Challenges in mobile WSN include deployment, localization, self-organization, navigation and control, coverage, energy, maintenance, and data proces**s**.

## 3 CHALLENGES OF POWER CONSUMPTION

Power consumption is a central design consideration for wireless sensor networks whether they are powered using batteries or energy harvesters. Vital to success for either approach, however, is the need for hardware that uses power intelligently. Following are the two approaches to minimize the power consumption of Wireless Sensor Networks.

### 3.1 Ultra Low-Power Networks

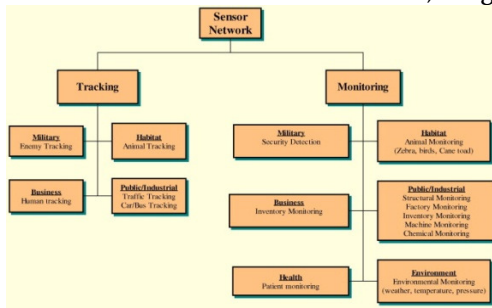

Fig. 1. Overview of Wireless Sensors Network

Several essential issues are key to developing low-power wireless sensor applications i.e. efficiently harvesting, converting, and storing the energy as well as using available energy in the most efficient way, without compromising performance (range, data rate, latency, and/or standards compliance).today this is becoming possible because of the development of ultra-low-power transceiver radio chip.

This chip usually works in combination with a microcontroller (MCU), which manages the transceiver ; switching it on, making it listen, transmit, wait for or receive acknowledge signals, or re-transmit, all in accordance to the communications protocol being used. Handling the transceiver's communication protocol activities, such as listening to whether the medium is free or waiting for an acknowledgement that a transmission has been received, requires the MCU to be awake the entire time and therefore consuming power. With a communication controller, the transceiver can transmit and receive the data independently from the MCU so that the MCU is only awake and in use when further data processing is needed. By using a communication controller-centric chip design shown in Fig. 2 rather than a microcontroller-centric design, and by using synchronized wake-ups, it is possible to reduce overall power consumption by 65% or more by eliminating the use of the MCU for the transceiver management functions.

Fig.3 shows how, by letting the communications controller decide when to wake up and check for messages, it is possible to greatly reduce overall energy consumption.

### 3.2 Energy Harvesting Solutions: Bursters vs. Tricklers

Energy harvesting devices can usually be divided in to two general categories[6]:

- Bursters- Produces a short but strong spike of energy. Examples are dynamo or micro-generator. It is a miniature AC generator in which motion is used to create energy instead of energy being used to create motion. Another energy burster is the piezo element, where energy is created from mechanical torsion of a piezoelectric material, which produces an electrical charge in response to an applied physical stress.

- Tricklers-Power source with a weak but steady output shown in Fig. 4. Solar cell is the example. Certain varieties of solar cells can extract energy from the limited light that is available indoors. To be usable for sense and control networks, the energy must be stored and the usage of the energy controlled. Another example is Peltier element that uses temperature differences as its energy source of choice, e.g., a wall of a house, where the outside temperature differs from that of the inside. Usually a temperature difference of 5°C gives enough usable energy to accumulate for consumption.

There is plenty of power in the environment and there are multiple options available to tap into this in order to power autonomous sensors[7]. However, at this moment, there is no one single solution for all applications—each power system needs to be customized to its specific application and also, it may require simultaneous use of two or more energy harvesting technologies.

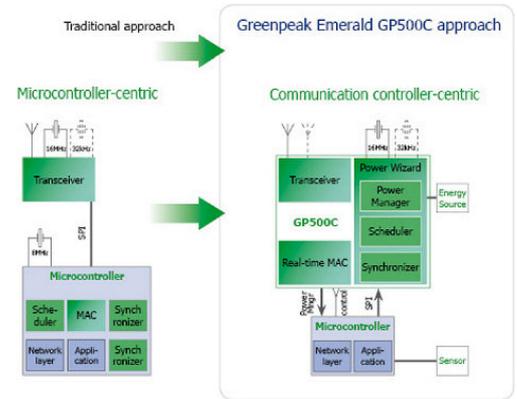

Fig. 2. Communication-controller-centric architecture vs.a traditional microcontroller-centric approach

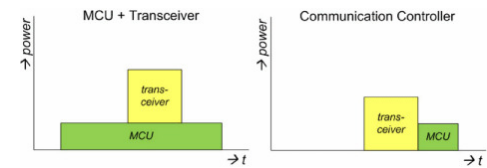

Fig. 3. By letting the microprocessor sleep until it is needed, it is possible to save >65% of energy usage as compared to the typical always-on transceiver

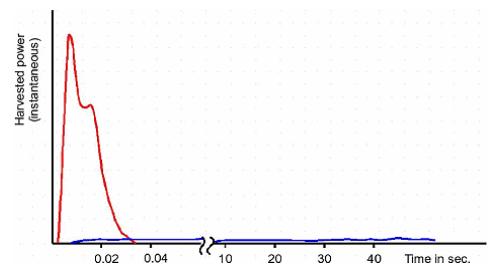

Fig. 4. Harvested power characteristics of a burster (red) and trickler (blue) showing the variation of available energy with time

## 4. INTELLIGENT BUILDING

Sensor networks will play a fundamental role in future intelligent buildings. There are two basic areas that we envision this technology having a critical impact, these are **energy conservation** and **security**. Most of the power consumption in most countries is done inside buildings for security and comfort living. Regarding security, it is clear that having a large number of sensors in the building that can monitor human movement further increase the levels of security in the building. [8]Having multiple sensing points can provide greater capability, as is leveraged commonly by sensor arrays such as in radar systems or binocular vision. Sensor networks inside the building should use small and non-intrusive devices. They have relatively low energy consumption at the system level, which implies long lifetime per deployment.

### 4.1 Challenges

The main challenges to move forward are:

**Hardware Cost**: The current cost of each individual sensor unit is still very high. Commercially available platforms cost in the order of Rs. 5000 per unit with temperature, humidity and light sensors when bought in large quantities. Capable sensors able to track human mobility inside buildings are costing around Rs.15000 per unit.

**System Architecture**: There is no unified system and networking architecture that is stable and mature enough to build different applications on top. Most of the applications and research prototypes are vertically integrated in order to maximize performance.

**Wireless Connectivity**: Wireless communication in indoor environments is still quite unpredictable using low-power consumption RF transceivers, in particular in clutter environments common inside buildings, with many interfering electromagnetic fields, such as the one produced by elevators, machinery and computers, among others.

**Programmability**: Some form of network re-programmability is desirable; doing so in energy and communication conservative form remains a challenge.

**Security**: The security challenges are at many levels.

- From the system point of view, it is critical that the information provided by the nodes be authenticated and the integrity verified, since this information provides the feedback loop to expensive equipment controlling power consumption in the building.
- From the users' point of view, it is also critical that this information cannot be easily spoofed and remains protected in the back end processor, since it may affect the privacy of users.

Each of the challenges above set the direction for many of the information technology research needs.

## 5. WIRELESS SENSOR NETWORKS FOR IN-HOME HEALTHCARE:

Advances in wireless sensor networking have opened up new opportunities in healthcare systems[9]. The future will see the integration of the abundance of existing specialized medical technology with pervasive, wireless networks. They will co-exist with the installed infrastructure, augmenting data collection and real-time response. Examples of areas in which future medical systems can benefit the most from wireless sensor networks are in-home assistance, smart nursing homes, and clinical trial and research augmentation.

In-home pervasive networks may assist residents by providing memory enhancement, control of home appliances, medical data lookup, and emergency communication.

Unobtrusive, wearable sensors will allow vast amounts of data to be collected and mined for next-generation clinical trials. Data will be collected and reported automatically, reducing the cost and inconvenience of regular visits to the physician. Therefore, many more study participants may be enrolled, benefiting biological, pharmaceutical, and medical-applications research.

### 5.1 Critical Development in Enabling Technologies for Future Medical Devices

- *Interoperability:* As a result of the heterogeneity present in the system, communication between devices may occupy multiple bands and use different protocols[10]. For example, motes use unlicensed bands for general telemetry or ISM equipment. Implanted medical devices may use a licensed band allocated for that purpose by the FCC. In order to avoid interference in the increasingly crowded unlicensed ISM band, biomedical devices may use the WMTS band (wireless medical telemetry services, at 608 MHz). [1] The homecare network must provide middleware interoperability between disparate devices, and support unique relationships among devices, such as implants and their outside controllers.

- *Real-time data acquisition and analysis:* The rate of collection of data is higher in this type of network than in many environmental studies. Efficient communication and processing will be essential. Event ordering, time-stamping, synchronization, and quick response in emergency situations will all be required.

- *Reliability and robustness:* Sensors and other devices must operate with enough reliability to yield high-confidence data suitable for medical diagnosis and treatment. Since the network will not be maintained in a controlled environment, devices must be robust.

- *New node architectures:* The integration of different types of sensors, RFID tags, and back-channel long-haul networks may necessitate new and modular node architectures.

## 6. ROADMAP: NEXT GENERATION SMART HOMECARE

Wireless Sensor Network architecture for Smart homecare that possesses the essential elements of each of the future medical applications are[11]:
- Integration with existing medical practices and technology,
- Real-time, long-term, remote monitoring,
- Miniature, wearable sensors, and
- Assistance to the elderly and chronic patients.

In smart homecare, the WSN collects data according to a Physician's specifications, removing some of the cognitive[12] burden from the patient (who may suffer age-related memory decline) and providing a continuous record to assist diagnosis. In-home tasks are also made easier, for example, remote device control, medicine reminders, object location, and emergency communication.

The architecture is multi-tiered, with both lightweight mobile components and more powerful stationary devices. Sensors are heterogeneous, and all integrate into the home network. Multiple patients and their resident family members are differentiated for sensing tasks and access privileges.

Examples of envisioned missions where the WSNs can quickly make an impact are the following[13]:

- *Sleep apnea.* Every night, monitor blood oxygenation, breathing, heart rate, EEG, and EOG using on-body sensors to assess severity and pattern of obstructive sleep apnea. Home network monitors agitation (movement) and stores and reports sensor data. Network alerts provider and patient if oxygenation falls below a threshold. Monitoring can continue while treatment efficacy is assessed

- *Journaling support.* Journaling is a technique recommended for patients to help their physicians diagnose ailments like rheumatic diseases. Patients record changes in body functions (range of motion, pain, fatigue, sleep, headache, irritability, etc), and attempt to correlate them with environmental, behavioral, or pharmaceutical changes. The homecare network can aid patients by: providing a time-synchronized channel for recording and transmitting the journal (PC, PDA, "dizziness" button); recording environmental data or external stimuli (temperature, barometric pressure, sunlight exposure, medication schedule); and quantitatively measuring changes in symptoms (pain, heart-rate, sleep disruption).

- Cardiac health. Cardiac arrhythmia is any change from the normal beating of the heart. Abnormal heart rhythms can cause the heart to be less efficient, and can cause symptoms such as dizziness, fainting, or fatigue. Since they are sometimes very brief, it can be difficult to properly characterize them. Cardiac stress tests attempt to induce the event while the patient is wearing sensors in a laboratory. In a homecare setting, wearable EKG sensors can monitor for the condition continuously, over days or weeks, until the event occurs. The recorded data is promptly sent to the physician for analysis.

## 7 WIRELESS SENSOR NETWORKS IN AGRICULTURE

WSN can be deployed for getting the information regarding soil degradation and water scarcity .[14]The IFDC, an international centre for soil fertility and agricultural development, reported that owing to the limitations in farming practices such as fertiliser usage, the levels of soil nutrients are declining at an annual rate of 30 Kg /ha in 85 % of African farm land [15]. Growers must then cultivate more land at the demise of wildlife and forest. This cycle of degradation was confirmed by IFDC researchers who further reported that agriculture, in conjunction with factors such as deforestation, worsen soil erosion and that if erosion rates continue unabated, the yield of some crops could fall by 1730% by 2020. This suggests the need for monitoring using a WSN to provide a basis for policy action.

Limited fresh water Agricultural production accounts for 70% of the world's fresh water usage and as growing population increases agricultural demands, per capita fresh water availability is declining faster than it is being replenished. This is compounded by the fact that irrigation relies heavily on ground water. For example 60% of irrigation water in Texas and 38% in California is pumped from the ground [16]. This heavy reliance on ground water for irrigation also exists in developing countries including Africa An estimated 16% of the world's crops are irrigated and 33% of the world's food is produced on irrigated land. In addition, the predicted global warming could increase world irrigation requirements by 26% to maintain current production rates [16]. This pressure on the world's water resources makes WSNs desirable and urgent.

Deployment of wireless sensor networks in agriculture is at its infancy. Currently three main wireless standards are used namely: WiFi, Bluetooth and ZigBee. Of these, ZigBee is the most promising standard owing to its low power consumption and simple networking configuration. However ZigBee standardisation is not yet complete. Some of the main obstacles in WSNs inspired by Ning Wang et al . [14]

The reliability of wireless sensors in agriculture is unproven and is considered risky and has many challenges [4].Obviously, no unique solution for all the challenges exist however the application of wireless sensors in land management can raise awareness of the effectiveness of new technologies in the agricultural domain.

## 6. CONCLUSION

Despite the technological obstacles such as the incomplete standardization of Zig Bee, and too much energy constraints the application of wireless sensors are tremendous. The paper concludes by considering its application in Intelligent Building, Smart Home Assistance and in agriculture .It has the potential to be an economically viable replacement to wired networks. They can provide risk assessment data such as alerting farmers at the onset of frost damage and providing better microclimate awareness, say, through the use of local temperature highs.

There is a credible future in the deployment of self organizing wireless sensors in agriculture, Medical and in designing low cost secured intelligent building for developing countries. The next stage of this research will investigate which parts of sensor technology are applicable and useful to decision makers in the land management domain.

**Dr Neelam Srivastava** She Graduated from M.M.M Engineering College, Gorakhpur in the year 1985 in Electronics and M.Tech in Microwave Engineering from Banaras Hindu University, Varanasi in the year 1987. She completed her PhD in the field of Optical communication from IET, Faculty of Engineering & Technology, Lucknow University. She is presently working as Assistant professor in Electronics Engineering Department of Institute of Engineering Technology, Lucknow since 1986. She is having a teaching Experience of around 24 years at UG and PG level. She uses to teach subjects related to various field of communications.. Presently her research area is wireless communication. She wrote Books in the field of Microwaves Communication and Telecommunication Switching. She is at present reviewer of many international journals. She is Fellow of IETE and Member IEE Societies.